\documentclass[aps,twocolumn,showpacs]{revtex4}
\usepackage{graphicx}
\usepackage{epsf}
\usepackage{amsmath}
\usepackage{textcomp}
\usepackage{amssymb}

\begin{document}

\title{Relativistic regimes for dispersive shock-waves in non-paraxial nonlinear optics}
\vspace{0.5cm}

\author{Silvia Gentilini$^{1\star}$, Eugenio DelRe$^2$, Claudio Conti$^{1,2}$}
\affiliation{\small
$^1$ ISC-CNR c/o Dipartimento di Fisica - Universit\`{a} La Sapienza, P. A. Moro
2, 00185, Roma, Italy\\
{$^2$ Dipartimento di Fisica - Universit\`{a} La Sapienza, P. A. Moro
2, 00185, Roma, Italy\\
$^\star$Corresponding author : silvia.gentilini@roma1.infn.it}}

\date{\today}

\begin{abstract}
We investigate the effect of non-paraxiality in the dynamics
of dispersive shock waves in the defocusing nonlinear Schr\"{o}dinger equation.
We show that the problem can be described in terms of a relativistic particle
moving in a potential. Lowest order corrections enhance the
wave-breaking and impose a limit to the highest achievable spectrum in an amount
experimentally testable.
\end{abstract}


\maketitle
\section{Introduction}
Dispersive shock waves (DSWs) have been the subject of intense research in the field of nonlinear waves, with specific applications in Bose-Einstein condensation  \cite{hoefer2006} and nonlinear optics  \cite{Barsi2007}, and are part of the large number of hydrodynamic-like phenomena  \cite{fleisher2012} that are considered important because of their links with quantum fluids \cite{Carusotto2013}, turbulence \cite{bortolozzo09, picozzi10}, disordered and curved systems \cite{Ghofraniha2012,Conti2013c}, and their application to laser physics \cite{Grelu14}.

With specific reference to nonlinear optics, much effort has been devoted to the formation of the DSWs. However,
all the reported theoretical investigations in the spatial domain are based on the paraxial approximation of the propagation equation of the electromagnetic field, within the validity of the hydrodynamical approach.
Shock waves are highly nonlinear processes that induce a substantial amount
of spectral broadening. We hence expect that non-paraxial terms are relevant in the development of the wave-breaking phenomena. An open issue is the identification of possible experimental
signatures of these effects. Indeed, non-paraxiality was previously investigated in the formation of solitons \cite{KivsharBook,Feit:88,Ciattoni:02,Kolesik2004a,Conti05PRL, Baruch:08,Alberucci:11}, however, so far, non-paraxial DSWs have not been considered.

In this manuscript we investigate theoretically and numerically the effect of non-paraxiality in the DSWs, and find that it limits the broadening of the spatial spectrum in the very same way special relativity limits the achievable velocities for a massive particle in the presence of a conservative force. In recent literature various authors have outlined analogies between nonlinear optics and relativistic regimes in the specific case of pulse propagation \cite{Philbin:08,Belgiorno:10,McDonald:10,Smolyaninov:13}. Here we extend the analogy to the phenomenon of the DSWs also in the spatial domain.

This manuscript is organized as follows. In section II, we review the leading model and the derivation of the hydrodynamic limit. In section III, we describe the link with relativistic dynamics and the way this link allows to predict the maximum phase gradient (velocity) and the shock point in the one-dimensional (1D) case. In section IV, we report the numerical simulations of the leading model in the two-dimensional (2D) case. Conclusions are drawn in section V.

\section{Model}
At the lowest order of perturbation the non-paraxial correction to the Foch-Leontovich equation for a paraxial beam described by a complex envelope $A$,
normalized such that $|A|^2$ is the optical intensity, can be written as \cite{Conti05PRL}:
\begin{equation}
i\frac{\partial A}{\partial Z}+\biggl(\frac{\nabla_{X,Y}^2}{2k}-\frac{\nabla_{X,Y}^4}{8k^3}\biggr)A+k\frac{ n_2}{n_0}|A|^2A=0\text{,}
\label{eq:notparaxial2}\end{equation}
letting $\lambda$ be the wavelength,
$k=2\pi n_0/\lambda$ the wavenumber, $n_0$ the bulk refractive index, and taking a nonlinear Kerr medium, with refracting index perturbation $n_2 |A|^2$.
In Eq. (\ref{eq:notparaxial2}) we neglect vectorial corrections to the nonlinear term \cite{Ciattoni00,Ciattoni:02}, as we make reference to highly nonlinear processes, such as thermal
effects and electrostrictive nonlinearity, for which vectorial effects are known to be negligible \cite{BoydBook,ghofraniha07}.

We consider the evolution of a focused Gaussian beam with profile at $Z=0$, $A=\sqrt{I_0}\exp(-X^2/4 w_0^2-Y^2/4 w_0^2)$ where $w_0$ is the beam waist. By introducing the scaled coordinates $(x',y',z')=(X/w_0,Y/w_0,Z/L_d)$,  with $L_d=kw_0^2$ the diffraction length, and the normalized variable $\psi=A/\sqrt{I_0}$, Eq. (\ref{eq:notparaxial2}) can be conveniently rewritten as follows, when $n_2<0$ :
\begin{equation}
i\frac{\partial\psi}{\partial z'}+\biggr[\frac{(\nabla'_{\bot})^2}{2}-\varepsilon\frac{(\nabla'_{\bot})^4}{8}\biggl]\psi-|\psi|^2\psi=0\text{,}
\label{eq:notparaxial3}\end{equation}
where $(\nabla'_{\bot})^2=(\frac{\partial^2}{\partial x'^2}+\frac{\partial^2}{\partial y'^2})$, $\varepsilon=\frac{1}{kL_d}=\frac{\lambda^2}{4\pi^2w_0^2}$, having chosen $I_0=\frac{n_0}{kL_d|n_2|}$.

By writing the normalized field as $\psi(r',z')=\sqrt{\rho(r',z')}\exp[i\phi(r',z')]$, where $r'=\sqrt{x'^2+y'^2}$, we can study Eq. (\ref{eq:notparaxial3}) in the framework of the WKB approximation \cite{bronski1994,forest98,kodama1999}.

In order to resort to the hydrodynamic approximation we introduce a small scaling factor $\eta$ such that $\psi(r,z)=\sqrt{\rho(r,z)}\exp[i\phi(r,z)/\eta]$, $z\rightarrow z'/\eta$, and $(x,y)\rightarrow(x'/\eta,y'/\eta)$; substituting in Eq. (\ref{eq:notparaxial3}) we obtain:
\begin{equation}
i\eta\psi_z+\frac{\eta^2}{2}\nabla_{\bot}^2\psi-\varepsilon\frac{\eta^4}{8}\nabla^4_{\bot}\psi-|\psi|^2\psi=0
\label{eq:notparaxial_WKB}
\end{equation}
where $\nabla_{\bot}^2=(\frac{\partial^2}{\partial x^2}+\frac{\partial^2}{\partial y^2})$.

Simple analytical treatment of the problem under consideration can be done in hydrodynamical approximation in 1D case as detailed in the following. As we are interested to the experimentally relevant 2D case, we compare in a later section the following theoretical results with 2D numerical simulations of Eq. (\ref{eq:notparaxial3}).

At the lowest order in $\eta$, the hydrodynamical approximation prescribes a density $\rho=\rho(x)$ independent by the propagation direction $z$, hence Eq. (\ref{eq:notparaxial_WKB}) reduces to the following equation for the phase $\phi$:
\begin{equation}
\phi_z+\frac{1}{2}\phi_x^2+\frac{\varepsilon}{8}\phi_x^4=-\rho(x)\text{.}
\label{eq:notparax_WKBredox}\end{equation}
By defining a velocity field as $v=\phi_x$ and differentiating w.r.t. $z$, we obtain the following equation:
\begin{equation}
v_z+vv_x+\frac{\varepsilon}{2}v^3v_x=-\partial_x\rho(x).
\label{eq:phase_notparaxial}\end{equation}
We notice that Eq. (\ref{eq:phase_notparaxial}) is formally similar to the Hopf equation, the solutions of which are known to develop the wave-breaking phenomenon \cite{WhithamBook}.
Non-paraxiality induces the higher order term $v^3v_x$.
\section{Link with relativistic dynamics}
The effect of the non-paraxiality on the shock point is determined in the following by the method of characteristic lines \cite{WhithamBook}, which allows us to express the solution of Eq. (\ref{eq:phase_notparaxial}) in terms of Hamiltonian system of ordinary differential equations:
\begin{equation}
\begin{array}{l}
\displaystyle \frac{d v}{d z}=f(x)=-\partial_x \rho(x)=-\frac{\partial H(x,v)}{\partial x}\\
\displaystyle \frac{d x}{d z}=v+\frac{\varepsilon}{2}v^3=\frac{\partial H(x,v)}{\partial v}\text{,}
\label{eq:characteristics}
\end{array}
\end{equation}
where $H(x,v)=\frac{v^2}{2}+\frac{\varepsilon}{8}v^4+\rho(x)$ is the conserved Hamiltonian, and $f(x)=-\partial_x\rho(x)$, is a conservative force,
with the intensity profile $\rho(x)$ playing the role of the potential. The non-paraxial term, weighted by $\varepsilon$, gives a contribution
which resembles the relativistic correction to the motion of a particle.
In fact, in special relativity the dynamics of a single particle subject to a conservative force $-\partial_x \rho(x)$ with rest mass, $m_0$, is given by the Hamiltonian $H_{RL}(x,v)=\gamma c^2 +\rho(x)$ with the Lorentz factor $\gamma=1/\sqrt{1-v^2/c^2}$ and $c$ the velocity of light. In the limit $v<<c$, $H_{RL}(x,v)\simeq m_0 c^2+\frac{m_0 v^2}{2}+\frac{3 m_0 v^4}{8 c^2}+\rho(x)$; in units such that $m_0=1$, this gives the Hamiltonian dynamics (\ref{eq:characteristics}) with $\varepsilon=3/c^2$.
\subsection{Maximal velocity}
\begin{figure}
\centerline{\includegraphics[width=\columnwidth]{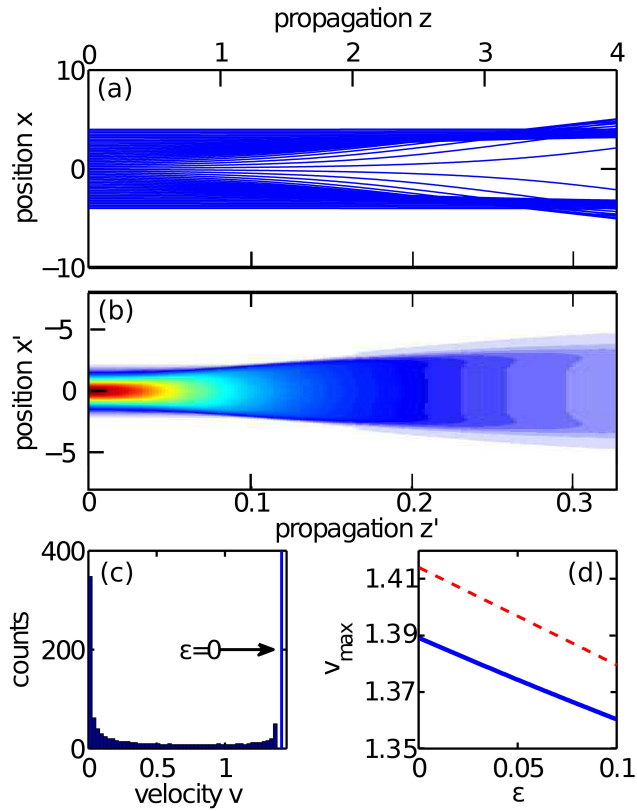}}
\caption{(a) Characteristic lines after Eq. (\ref{eq:characteristics}) versus $z$ for $\varepsilon=0.01$ and $\rho(x)=\exp(-x^2/2)$; (b) intensity map in the plane $(x',z')$ as obtained by the 2D-BPM simulations of Eq. (\ref{eq:notparaxial3}) in correspondence of the excitation of DSW with power $P=130$mW, waist $w_0=1\mu$m, and $\varepsilon=0.01$;
(c) histogram of the distribution of velocities $v$ at $z=4$ for
$\varepsilon=0.01$, the vertical line showing the $v_{max}$ for $\varepsilon=0$; (d) $v_{max}$ vs $\varepsilon$ as obtained by numerical integration of Eq. (\ref{eq:characteristics}) (continuous line) and by plotting Eq. (\ref{vmax}) (dashed line)}\label{figtheory1}
\end{figure}

The analogy with the relativistic dynamics indicates that effect of non-paraxiality reduces the spatial spectrum resulting from the shock.
Indeed the velocity $v$ of the effective particle corresponds to wavector $k$ of an optical ray.
By using the conservation of $H(x,v)$ in (\ref{eq:characteristics}),
the case of an input beam with a flat phase front, corresponds to an initial distribution particles with zero velocity positioned in a potential $\rho(x)$ given by the intensity profile of the beam. Hence, at $z=0$ all the particles have a distribution of potential energy $\rho(x)$ that, upon propagation, is converted in kinetic energy. The condition $H(x,v)=v^2/2+\varepsilon v^4/8=\rho(x)$ gives the maximal velocity $v_{MAX}$ of a characteristic line originally placed in $x$.
For the considered Gaussian beam $\rho(x)=\exp(-x^2/2)$, the particles located in proximity of the peak intensity $x=0$ have the
highest velocity and collide upon propagation with those located at the beam edges causing the hydrodynamic shock (see Fig. \ref{figtheory1}(a)).
The conservation of $H(x,v)$ shows that $v_{MAX}$ is reduced when increasing $\varepsilon$:
\begin{equation}
v_{MAX}(\varepsilon)=v_{MAX}(0)\left(1-\frac{\varepsilon}{4}\right)+O(\varepsilon^2)\text{,}
\label{vmax}
\end{equation}
with $v_{MAX}(0)=\sqrt{2}$ in the Gaussian case $\rho(x)=\exp(-x^2/2)$. Equation (\ref{vmax}) predicts that after the shock, non-paraxial effects limit the maximal achievable velocity. The distribution of velocity is directly measurable by the far-field in optical measurements \cite{Gentilini:12}.

%--------------Figure 1------------------------%
%----------------end figure 1-----------------%
\subsection{Shock point}
As shown in Fig. \ref{figtheory1}(a) the shock is signaled by the caustic resulting from the envelope of characteristic lines at the boundary of the beam.
The lines in these regions are parabolic, and starting from a point $x_0$ with $v=0$, they can be analytically approximated by solving Eq. (\ref{eq:characteristics}) with $f(x)=f(x_0)$ approximately constant, which gives an estimate of the shock point $z_S$. Considering two infinitesimally near characteristic lines starting at $x_0+dx/2$ and $x_0-dx/2$ indicated respectively
as $x_+(z)$ and $x_-(z)$, the shock point can be found by the condition $x_+(z)=x_-(z)$.
Eqs. (\ref{eq:characteristics}) can be solved by direct integration by
taking $f(x)=f(x_0\pm dx/2)$, and we have ($x_{0\pm}\equiv x_0\pm dx/2$):
\begin{equation}
x_\pm(z)=f(x_{0\pm})\frac{z^2}{2}+\frac{\varepsilon}{8}f(x_{0\pm})^3 z^4+x_{0\pm}\text{.}
\end{equation}
In previous work \cite{Ghofraniha2012}, the numerical solutions for $\varepsilon=0$ have been found,
but so far no analytical evaluation of the shock point has been given. Denoting $z_{0s}$ the shock point in the paraxial case, we find
\begin{equation}
z_{0}^2=-\frac{2}{f(x_0)'}=\frac{2}{|\rho(x_0)''|}
\label{shockpointparaxial}
\end{equation}
which shows that the shock occurs in the regions where $\rho(x_0)''<0$, the prime denoting the differentiation w.r.t. $x$. Considering the Gaussian case, the minimum of $z_0$ in Eq. (\ref{shockpointparaxial}) w.r.t. $x_0$ gives the paraxial shock point, i.e., $x_0=\sqrt{3}$ and $z_{0s}=\exp(3/4)\cong 2.1$, which is in quantitative agreement with numerical simulations previously reported (as, e.g., in \cite{ghofraniha07}).

In the non-paraxial case, we use $\varepsilon$ as a perturbation parameter and the shock condition $x_+=x_-$ gives:
\begin{equation}
3\varepsilon f(x)^2 f(x)' z_1^4+4f'(x) z_1^2+8=0
\end{equation}
which solved w.r.t. $z_1^2$ gives:
\begin{equation}
z_1^2=-\frac{2}{f'}-\frac{3\varepsilon f^2}{(f')^2}=z_0^2-\frac{3\varepsilon f^2}{(f')^2}
\label{eq:z1}\end{equation}
by calculating the derivatives $f=-\rho_x$ and $f'=-\rho_{xx}$ and substituting in Eq. (\ref{eq:notparaxial3}), this latter can be rewritten as:
\begin{equation}
z^2_1=\frac{2e^{x^2/2}}{x^2-1}-3\varepsilon\frac{x^2}{(x^2-1)^2}\text{.}
\end{equation}
In order to calculate the minimum w.r.t. $x$, we calculate the derivative of $z_1^2$ and solve the following equation:
\begin{equation}
\frac{dz_1^2}{dx}=\frac{2x}{(x^2-1)^2}[e^{x^2/2}(x^4-4x^2+3)+3\varepsilon(x^2+1)]=0,
\label{eq:derivativez1}\end{equation}
by performing the variable change $x=\sqrt{3}+\varepsilon x_1$ and retaining only the leading order in $\varepsilon$, Eq. (\ref{eq:derivativez1}) reduces to:
\begin{equation}
\frac{dz_1^2}{dx}=\varepsilon(12+4\sqrt{3}e^{3/2}x_1)=0
\end{equation}
which gives the minimizing value for $x$, $x_{MIN}=\sqrt{3}(1-0.2\varepsilon)$. Substituting such value in Eq. (\ref{eq:z1}) and retaining only the leading terms in $\varepsilon$, we obtain:
\begin{equation}
z_1=e^{3/4}(1-\frac{9}{8}\varepsilon e^{-3/2})=z_{0s}(1-0.25\varepsilon)+O(\varepsilon^2).
\label{shockpoint}
\end{equation}
This shows that the shock point is anticipated by the non-paraxial terms.

By the above theoretical analysis emerges that in the non paraxial regime the diffraction is much more enhanced and hence the nonlinear effect producing the spectral broadening is limited w.r.t. the paraxial case. On the other hand the enhanced diffraction favours the collisions of the characteristic lines (direction of the propagation energy) and anticipate the position of the shock point.

\section{Numerical simulations}

%----------figure 2------------------------------------%
\begin{figure*}
\centerline{\includegraphics[width=\textwidth]{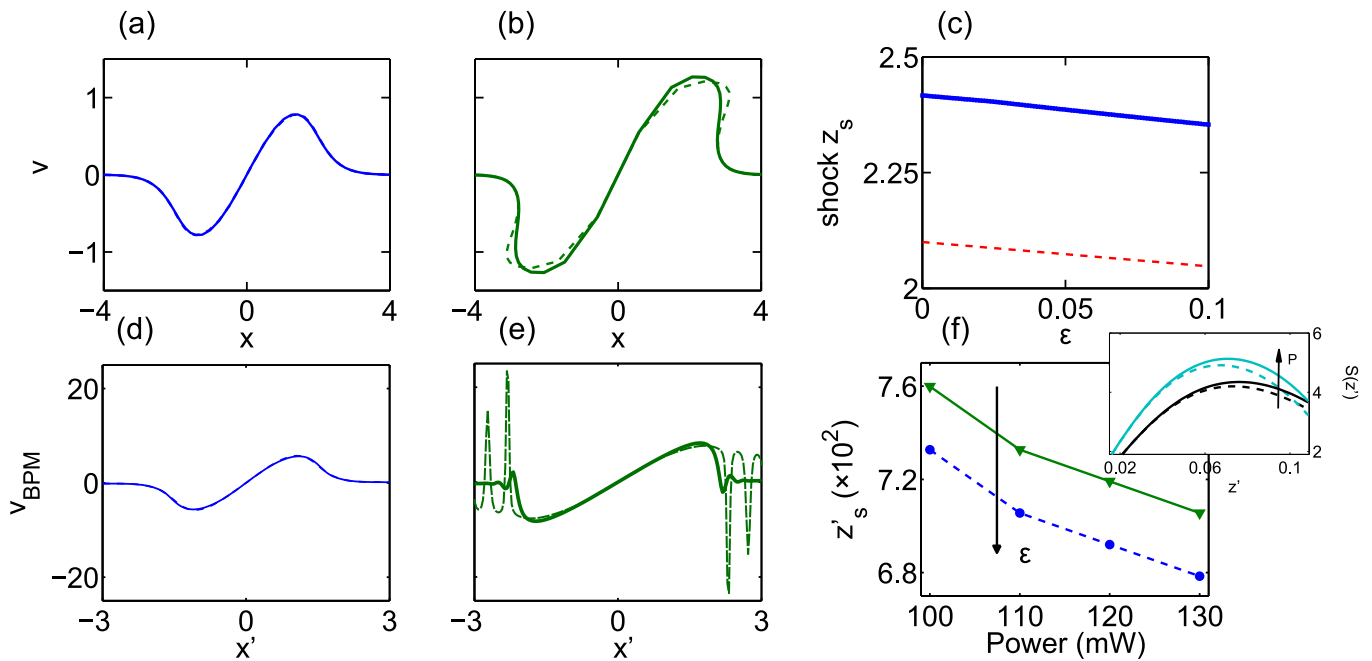}}
\caption{(a-b) Velocity profile versus position after Eq. (\ref{eq:characteristics}) showing the folding of the wave at the shock for $\varepsilon=0$ (continuous line) $\varepsilon=0.5$ (dashed line) for $z=1.3$ (a) and $z=2.7$ (b). Notice that the two lines in panel (a) cannot be distinguished; (c) $z_S$ vs $\varepsilon$ calculated as the point of maximal $v_x$ from the characteristic lines (continuous line) and from Eq. (\ref{shockpoint}) (dashed line). (d)-(e) The same of panels (a) and (b) obtained by solving the Eq.(2) with the BPM at two different propagation distances: $z'=0.05$ (d) and $z'=0.1$ (e);(f) shock point $z'_s$ vs $P$ calculated as the point of maximal phase steepness for $\varepsilon=0$ (continuous line) and $\varepsilon=0.03$ (dashed line). The inset shows the phase steepness, S, vs the propagation direction $z'$ for different $P$ and for $\varepsilon=0$ (continuous lines) and $\varepsilon=0.01$ (dashed line)}\label{figtheory2}
\end{figure*}
%-----------end figure2----------------------------------%
In the following we compare the above 1D theoretical analysis with the numerical solutions of Eq. (\ref{eq:characteristics}) obtained with the characteristics for $\rho(x)=\exp(-x^2/2)$; the corresponding trajectories for $\varepsilon=0.01$ are shown in Fig. \ref{figtheory1}(a).

We also test the 1D theory by resorting to 2D beam propagation method (BPM) to simulate Eq. (\ref{eq:notparaxial3}) by considering the propagation of a 2D Gaussian beam along the $z'$ direction in a de-focusing medium ($n_2<04$). In Fig. \ref{figtheory1}(b) is shown the intensity profile in the (x',z') plane as obtained by the BPM simulations in correspondence of a DSW excitation with the aim to provide a direct comparison with Eq. (\ref{eq:notparaxial3}).

The characteristic lines allow to retrieve the  histogram of the velocity distribution after the occurrence of the shock at $z=4$ as shown in Fig. \ref{figtheory1}(c). The large number of lines with zero velocity corresponds to trajectories located far away the peak intensity of the beam. Lines with velocity in the range $(0,v_{MAX})$ coincide with the tails of the beam. The many lines with maximal velocity ($v=1.4$) are related to the regions with high intensity. The maximal velocity in the paraxial case ($\varepsilon=0$) is indicated by the vertical line and reveals that the relativistic-like effect limits the maximally achievable velocity $v_{MAX}$, which is shown to linearly decrease with $\varepsilon$ (Fig. \ref{figtheory1}(d)), as predicted by Eq. (\ref{vmax}) (dashed line in Fig. \ref{figtheory1}(d)).

In Fig. \ref{figtheory2}(a) and Fig. \ref{figtheory2}(b), we show the velocity profiles vs position $x$ in the paraxial ($\varepsilon=0$, continuous line)
and non-paraxial case ($\varepsilon=0.5$, dashed line) at $z=1.3$ (a) and $z=2.7$ (b).
Note that the folding in the non-paraxial case appears more pronounced; this results in an anticipated shock point, as numerically
calculated from the characteristic lines and shown in  Fig. \ref{figtheory2}(c) (continuous line), which follows Eq. (\ref{shockpoint}) (dashed line). The discrepancy between the theoretical (dashed lines) and the numerical (continuous line) curves of Fig. \ref{figtheory1}(d) and Fig. \ref{figtheory2}(c) are due to the several approximations underlying Eq. (\ref{eq:characteristics}).

The comparison with the results obtained by the BPM simulations is provided by Fig. \ref{figtheory2}(d) and Fig. \ref{figtheory2}(e), where we show the velocity profiles, $v_{BPM}$, calculated as the derivative w.r.t. $x'$ of the transverse phase (i.e. the chirp, $d\phi/dx'$), for the paraxial (continuous line) and not paraxial (dashed line) for the two propagation distances $z'=0.05$ (d) and $z'=0.1$ (e). We notice that in the BPM simulations, after the shock, the \emph{undular bores} are clearly visible and regularize the occurrence of the singularity (see Fig. \ref{figtheory2}(e)). We recall that such a mechanism is not present in the Hopf equation (and hence in the characteristic lines), but arises from the diffraction terms present in Eq. (\ref{eq:notparaxial3}). We also notice that the \emph{undular bores} are more pronounced in the non paraxial case.

Figure \ref{figtheory2}(f) shows $z'_s$ vs $P$ in the paraxial (continuous line) and non-paraxial (dashed line) regime as obtained by the 2D BPM simulations; notably enough in the latter case the shock is anticipated when
$\varepsilon>0$ as predicted in the analysis above. We determine the shock point, $z'_s$, by calculating the transverse
phase at any value of the propagation coordinate $z'$ and determining the point at which its derivative w.r.t. $x'$, i.e. the chirp, is maximum. We denote the maximum chirp w.r.t. $x'$ as the ``steepness'', $S$. We show $S$ vs $z'$ in the inset of Fig. \ref{figtheory2}(f): the shock point, $z'_s$, is the point of maximal steepness.

%--------------Figure 3------------------------%
\begin{figure}
\centerline{\includegraphics[width=\columnwidth]{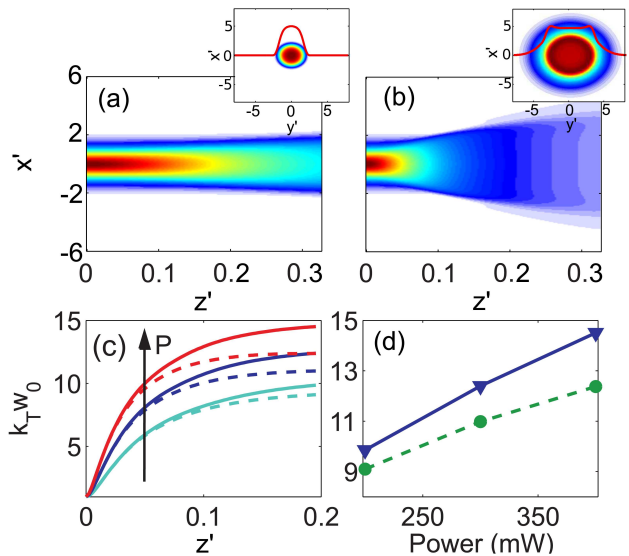}}
\caption{(a,b) Intensity map in the plane $(x',z')$ in the nonparaxial case ($\varepsilon=0.01$) as obtained by using an input waist beam of $w_0=1\mu$m for two different laser power, $P=10$mW(a), and $P=130$mW (b). The insets show the corresponding intensity profiles as they are retrieved at the output (x',y') plane; (c) spectral widths vs propagation length for $\varepsilon=0$ (continuous lines) and $\varepsilon=0.01$ (dashed lines) and for different powers $P$ (from bottom to top are $P=200,300,400$mW); (d) spectral width at $z'=0.2$ vs $P$ for $\varepsilon=0$ (continuous line) and for $\varepsilon=0.01$ (dashed line).}
\label{fignum1}\end{figure}
%----------------end figure 3-----------------%
%--------------Figure 4------------------------%
\begin{figure}
\centerline{\includegraphics[width=\columnwidth]{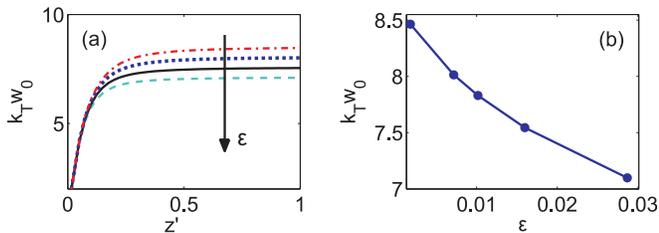}}
\caption{(a) Transverse spectral width $k_T w_0$ vs propagation coordinate
for different $\varepsilon$ (corresponding to $w_0$ in the range $0.5$ and
$2\mu$m) at fixed power $P=130mW$; (b) spectral width at $z'=1$ and $P=130$mW vs $\varepsilon$.}
\label{fignum2}\end{figure}
%----------------end figure 2-----------------%
To show that non paraxial corrections are relevant in the laboratory experiments, we repeat the 2D-BPM simulations of Eq. (\ref{eq:notparaxial3}) for several beam waists, $w_0$, ranging between $0.5 \mu$m and $2 \mu$m,
with wavelength $\lambda=532$nm, corresponding to values for $\varepsilon$ in the range $0.001$ and $0.03$.
For each waist, we change the input laser power, $P$, in the interval [$10$mW, $400$mW] by assuming $n_2=2\times10-12$W/m$^2$ following experimental investigations in aqueous solutions \cite{ghofraniha07,Gentilini2013}.

Figure \ref{fignum1}(a) and Fig. \ref{fignum1}(b) show the intensity distribution in the plane ($x'$,$z'$) as obtained from 2D propagation simulations of Eq. (\ref{eq:notparaxial3}) for a nonparaxial degree of $\varepsilon=0.01$, an input waist beam of $w_0=1\mu$m, at two different powers, $P=10$ and $130$mW respectively. The insets show the transverse intensity distribution as it appears at the output ($x',y'$) plane. At low power ($P=10$mW), we recognize a linear propagation regime dominated by diffraction; conversely at high power ($P=130mW$), we identify the onset of a nonlinear regime in the enhancement of the angular spreading along the beam propagation and in the emergence of an intense ring, characteristic of the DSWs, clearly visible in the intensity distribution at the ($x',y'$) output plane reported in the corresponding inset.

In Fig. \ref{fignum1}(c) is shown as for various $P$, the optical
spatial spectrum reaches a nearly steady width vs $z'$, after the shock generation, which can be quantified by the dimensionless parameter $k_T w_0$, being $k_T$ the transverse spectral width (calculated as the standard deviation of the spatial spectrum of the beam); when $\varepsilon\neq0$ (dashed lines in Fig. \ref{fignum1}(c)) the spectral width is limited. This is effect is also shown in Fig. \ref{fignum1}(d), where we show that output spectral width at $z'=0.2$ for various powers $P$ in the paraxial (continuous line) and non-paraxial (dashed line) cases. We stress that the wave breaking is clearly observable in all the considered cases and that, in agreement with the theoretical analysis above, the angular aperture of the laser beam is more pronounced in correspondence of the smaller waists.

We show in Fig. \ref{fignum2}(a) the spectral width vs propagation direction $z'$ for various values of $\varepsilon$, corresponding to various input waists $w_0$; in Fig. \ref{fignum2}(b) the output
spectral width at a fixed $z'=1$ is given vs $\varepsilon$, indicating the bandwidth limitation arising because of non-paraxiality.

\section{Conclusions}

We investigated the role of the paraxial approximation on the occurrence of the DSWs phenomenon.
By a theoretical approach based on the method of characteristics
we found that the problem can be analyzed in terms of the evolution
of relativistic particles; the effect of non-paraxiality is
on one hand enhancing the shock point and on the other hand limiting the
maximally achievable velocity, i.e., the spatial spectrum.
While it is mathematically well known that the nonlinear Schr\"{o}dinger equation supports blow-up solutions,
corresponding to analogues of singular shock waves in the hydrodynamic limit, the case of the higher order
derivatives that occurs beyond the paraxial limit,  as in Eq. (\ref{eq:notparaxial3}), is still open,
and our analysis shows that non-paraxiality does not prevent DSWs.
The relativistic terms in the resulting Hopf equation, Eq. (\ref{eq:phase_notparaxial}), limit
the angular aperture of the spatial beam during the shock by an amount that is experimentally accessible, as wave-simulations
of realistic experiments demonstrate.
The formal analogy with the propagation of a relativistic particle opens the road to a variety of further studies, such as considering properly designed wave-fronts to enhance the relativistic/non-paraxial regime, inducing collision of multiple-shocks, soliton generation, using incoherent beams to generate analogues of collisions of relativistic gases, as well as analyzing the role of disorder, random walk and diffusion in relativistic regimes.

\emph{Acknowledgments.} We acknowledge funding from the Italian Ministry of Research (MIUR) through the FIRB grant PHOCOS-RBFR08E7VA and through the PRIN project no. 2009P3K72Z and the project Sapienza Ricerca 2014, PhotoAnderson.


\begin{thebibliography}{44}

\bibitem{hoefer2006}
M.~A. Hoefer, M.~J. Ablowitz, I.~Coddington, E.~A. Cornell, P.~Engels, and
  V.~Schweikhard,
\newblock Phys. Rev. A \textbf{74}, 023623 (2006).

\bibitem{Barsi2007}
C. Barsi, W. Wan, C. Sun, and J.W. Fleischer,
\newblock Opt. Lett. \textbf{32}, 2930 (2007).

\bibitem{fleisher2012}
S. Jia, M. Haataja, and J.W. Fleischer,
\newblock New J. Phys. \textbf{14}, 075009 (2012).

\bibitem{Carusotto2013}
I. Carusotto and C. Ciuti,
\newblock Rev. Mod. Phys. \textbf{85}, 299 (2013).

\bibitem{bortolozzo09}
U. Bortolozzo, J. Laurie, S. Nazarenko, and S. Residori,
\newblock J. Opt. Soc. Am. B \textbf{26}, 2280 (2009).

\bibitem{picozzi10}
P. Suret, S. Randoux, H.R. Jauslin, and A. Picozzi,
\newblock Phys. Rev. Lett. \textbf{104}, 054101 (2010).

\bibitem{Ghofraniha2012}
N.~Ghofraniha, S.~Gentilini, V.~Folli, E.~DelRe, and C.~Conti,
\newblock Phys. Rev. Lett. \textbf{109}, 243902 (2012).

\bibitem{Conti2013c}
C.Conti,
\newblock Chin. Phys. Lett. \textbf{31}, 030501 (2014).

\bibitem{Grelu14}
C.Lecaplain, J.M. Soto-Crespo, Ph. Grelu, and C.Conti
\newblock Opt. Lett. \textbf{39}, 263 (2014).

\bibitem{KivsharBook}
Y.~Kivshar and G.~P. Agrawal,
\newblock {\em Optical solitons}
\newblock (Academic, New York, 2003).

\bibitem{Feit:88}
M.~D. Feit and Jr. J.~A.~Fleck,
\newblock J. Opt. Soc. Am. B \textbf{5}, 633 (1988).

\bibitem{Ciattoni:02}
A.~Ciattoni, C.~Conti, E.~DelRe, P.~Di Porto, B.~Crosignani, and A.~Yariv,
\newblock  Opt. Lett. \textbf{27}, 734 (2002).

\bibitem{Kolesik2004a}
M.~Kolesik and J.~V. Moloney,
\newblock Phys. Rev. E \textbf{70},036604 (2004).

\bibitem{Conti05PRL}
C.~Conti, G.~Ruocco, and S.~Trillo,
\newblock Phys. Rev. Lett. \textbf{95}, 183902 (2005).

\bibitem{Baruch:08}
G.~Baruch, G.~Fibich, and Semyon Tsynkov,
\newblock Opt. Express \textbf{16}, 13323 (2008).

\bibitem{Alberucci:11}
A. Alberucci and G. Assanto,
\newblock Opt. Lett. \textbf{36}, 193 (2011).

\bibitem{Philbin:08}
T.G.~Philbin, C. Kuklewicz, S. Robertson, S.Hill, F.K\"{o}nig, and U.Leonhardt,
\newblock Science \textbf{319}, 1367 (2008).

\bibitem{Belgiorno:10}
F.Belgiorno, S.L.Cacciatori, M.Clerici, V.Gorini, G.Ortenzi, L.Rizzi, E.Rubino, V.G.Sala, and D. Faccio,
\newblock  Phys. Rev. Lett. \textbf{105}, 203901 (2010).

\bibitem{McDonald:10}
G.S.McDonald, J.M. Christian, and T.F. Hodgkinson,
\newblock \lq\lq Optical soliton pulses with relativistic characteristics,\rq\rq
\newblock in {\em Proceedings of 5th International Conference on Advanced Optolelectronics andLasers}, (2010).

\bibitem{Smolyaninov:13}
I.I.Smolyaninov,
\newblock Phys. Rev. A \textbf{88}, 033843 (2013).

\bibitem{Ciattoni00}
A. Ciattoni, B. Crosignani, and P. Di Porto,
\newblock  Opt. Comm. \textbf{177}, 9 (2000).


\bibitem{BoydBook}
R.~W. Boyd,
\newblock {\em Nonlinear Optics}
\newblock (Academic, New York, 2002).

\bibitem{ghofraniha07}
N. Ghofraniha, C. Conti, G. Ruocco, and S. Trillo,
\newblock  Phys. Rev. Lett. \textbf{99}, 043903 (2007).

\bibitem{bronski1994}
J.~C. Bronski and D.W. McLaughlin,
\newblock {\em Singular Limits of Dispersive Waves}.
\newblock (Plenum, 1994).

\bibitem{forest98}
M.~G. Forest and K.T.-R. McLaughlin,
\newblock  J. Nonlinear Science \textbf{7}, 43 (1997).

\bibitem{kodama1999}
Y.Kodama,
\newblock  SIAM J. Appl. Math. \textbf{59},2162 (1999).

\bibitem{WhithamBook}
G.~B. Whitham,
\newblock {\em Linear and Nonlinear Waves},
\newblock (Wiley, New York, 1999).

\bibitem{Gentilini:12}
S.~Gentilini, N.~Ghofraniha, E.~DelRe, and C.~Conti,
\newblock Opt. Express \textbf{20}, 27369 (2012).

\bibitem{Gentilini2013}
S.~Gentilini, N.~Ghofraniha, E.~DelRe, and C.~Conti,
\newblock Phys. Rev. A \textbf{87}, 053811 (2013).

\end{thebibliography}
\end{document}